\documentclass{appolb}
\usepackage{epsfig}

\begin{document}
\title{The Long Slow Death of the HBT Puzzle
\thanks{Presented at Workshop on Particle Correlations and Femtoscopy, Krakow, 2008.\\Generous support from the U.S. Dept. of Energy through grant no. DE-FG02-03ER41259 is gratefully acknowledged.}%
}
\author{Scott Pratt
\address{Department of Physics and Astronomy,
Michigan State University\\
East Lansing, MI 48824~~USA}
}
\maketitle
\begin{abstract}
Femtoscopic measurements at RHIC have been hailed as a source of insight into the bulk properties of QCD matter. However, hydrodynamic models, which have been successful in reproducing other observables have failed to satisfactorily explain femtoscopic two-particle correlation measurements. This failure has been labeled the ``HBT puzzle''. In this talk, I present correlations using a hybrid model composed of a viscous hydrodynamic module and a hadronic cascade. After incorporating early acceleration, viscosity and a stiffer equation of state, the extracted source sizes come much closer to experiment. 
\end{abstract}
\PACS{25.75.-q, 25.75.Gz, 25.70.Pq}
  

At the dawn of the RHIC era, the most sophisticated, and seemingly most realistic, transport models failed miserably to match source sizes as inferred from experimental two-particle correlations measurements. These models \cite{Soff:2000eh,Teaney:2001av,Hirano:2002hv} applied a hydrodynamic code to model the super-hadronic stage, which was then coupled to a hadronic cascade whose purpose was to simulate the low density stage and breakup. Seemingly less realistic cascade models were not completely satisfactory, but performed significantly better \cite{Petersen:2008gy,Li:2008ge,Humanic:2006sk,Lin:2002gc}. These models ignored the partonic phase altogether, or at least ignored the softness of the equation of state associated with the phase transition. The failure of the hydrodynamic models picked up the name ``HBT puzzle'' (HBT refers to Hanbury-Brown and Twiss who developed interferometric source size measurements with photons \cite{HanburyBrown:1956pf}). A second feature of the puzzle concerned parametric fits, such as the blast-wave model, which could come close to the data but only with unphysically high breakup densities. Inspired by this puzzle, we have developed a viscous hydrodynamic model coupled to a hadronic cascade \cite{Pratt:2008sz}, and have studied the effect of several improvements or modifications of the initial treatments, and found that the femtoscopic data from RHIC can be well reproduced without either invoking any contentious ideas, or setting parameters outside usual expected values. Considered individually, none of these modifications explained more than half the original discrepancy. Instead several changes conspired to push source sizes in the same direction, mainly by making the reaction more  explosive which leads to smaller source sizes.
\begin{figure}
\centerline{\includegraphics[width=0.55\textwidth]{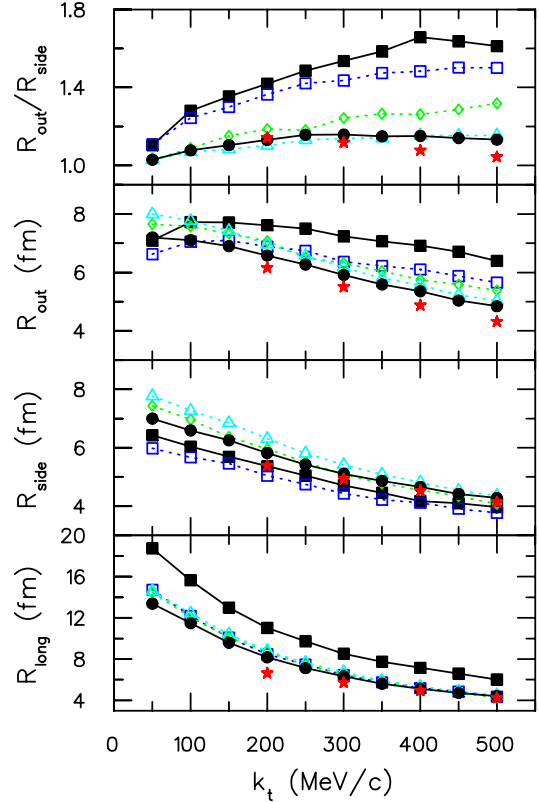}}
\caption{\label{fig:Rosl_everything}
Gaussian radii reflecting spatial sizes of outgoing phase space distributions in three directions: $R_{\rm out}$, $R_{\rm side}$ and $R_{\rm long}$. Data from the STAR collaboration (red stars) are poorly fit by a model with a first-order phase transition, no pre-thermal flow, and no viscosity (solid black squares). Correcting for all those deficiencies, and using a more appropriate treatment of the relative wave function in the Koonin equation \cite{Lisa:2005dd,Koonin:1977fh} brings calculations close to the data (filled black circles). The sequential effects of including pre-thermal acceleration (open blue squares), using a more realistic equation of state (open green diamonds), and adding viscosity (open cyan triangles) all make substantial improvements to fitting the data. An improved relative wave function yielded modest improvements (compare open cyan triangles to filled black circles).}
\end{figure}

Since many talks from this workshop have already described how femtoscopic source sizes can be extracted from data, and how they are theoretically related to the emission history of a heavy ion collision, I shall be brief, and mention a few key points. First, the three dimensions $R_{out},R_{\rm side}$ and $R_{\rm long}$ refer to the outward (parallel to the momentum of the particle), sideward (perpendicular to the beam and to the particle's momentum) and longitudinal (along the beam axis) directions. The sizes represent Gaussian fits to the spatial size and shape of the outgoing phase space cloud of particle's of a specific momentum. Since the initial fireball has little extent along the beam direction, the size $R_{\rm long}$ can be related to the time at which particles are created. If the emission is sudden, the two transverse dimensions tend to be similar, but for a longer duration of emission, particles emitted earlier get ahead and $R_{\rm out}$ becomes larger than $R_{\rm side}$. For central collisions, and for measurements at mid-rapidity, each dimension is a function of $p_t$ only. The aforementioned hydrodynamic models tended to over-predict $R_{\rm long}$ and the $R_{\rm out}/R_{\rm side}$ ratio, suggesting that the explosion was more rapid, and with a more sudden emission than that portrayed by the models.

The principle changes to the model responsible for improving the comparison with data are: pre-equilibrium flow, using a stiffer equation of state, and incorporating viscosity. A modest improvement also entails if one more realistically models the $\pi-\pi$ interaction. Additionally, we affirm the findings of \cite{Broniowski:2008vp} and see that the shape of the initial profile can significantly affect the source sizes. In order to illustrate the effects, Fig. \ref{fig:Rosl_everything} first presents a benchmark calculation, which has no pre-equilibrium flow, no viscosity and too stiff of an equation of state. The source sizes poorly match the data, and are similar to other models with the same lack of features. The $R_{\rm out}/R_{\rm side}$ ratio is approximately 40\% too high, and $R_{\rm long}$ is also over-predicted. By moving the start time of the calculation from 1.0 fm/$c$ to 0.1 fm/$c$, the collision becomes more explosive, which shortens the extent and duration of the pion emission, and leads to source sizes significantly closer to the data as shown in Fig. \ref{fig:Rosl_everything}. Similar effects have been seen in similar contexts in both hydrodynamic and microscopic models \cite{Gyulassy:2007zz,Li:2008ge,Broniowski:2008vp}. Pre-equilibrium flow is inevitable, as flow derives from the conservation of the stress-energy tensor, and its strength is largely independent of equilibration or the microscopic state of matter during the first fm/$c$ \cite{Vredevoogd:2008id}. 

The second change we consider is improving the equation of state. The equation of state used in the benchmark calculation was first order, with a large latent heat, $L=1.6$ GeV/fm$^3$, similar in spirit to what was used in previous studies \cite{Soff:2000eh,Teaney:2001av,Hirano:2002hv}. During the mixed phase, the speed of sound is zero and the pressure and temperature stay constant, which would lead to extended lifetimes, especially if the initially energy density were lower, and close to the maximum for the mixed phase \cite{Pratt:1986cc,Rischke:1996em}. Lattice calculations suggest a noticeably stiffer equation of state with no first-order transition. Figure \ref{fig:Rosl_everything} shows significantly improved source sizes from a calculation where the soft region is cut in half and the speed of sound, $c_s^2=dP/d\epsilon$, is set to 0.1$c$. As expected, the stiffer equation of state increases the explosivity and results in improved source sizes as illustrated in Fig. \ref{fig:Rosl_everything}.

The third modification to the hydrodynamic model is to include viscosity. The Navier-Stokes equation gives the following form for the  stress-energy tensor in the fluid rest frame,
\begin{equation}
T_{ij}=P\delta_{ij}-\eta\left\{\partial_iv_j+\partial_jv_i-(2/3)\nabla\cdot{\bf v}\delta_{ij}\right\}
-\zeta\nabla\cdot{\bf v},
\end{equation}
where $\eta$ and $\zeta$ are the shear and bulk viscosity coefficients respectively. In the early stages of a RHIC collision the velocity gradient is far stronger along the $z$ axis, and the shear correction is negative for $T_{zz}$ and positive for the transverse components of pressure, $T_{xx}$ and $T_{yy}$ (The correction for shear is traceless). This increases transverse acceleration which again significantly improves agreement with experimental source sizes as shown in Fig. \ref{fig:Rosl_everything}, and was also seen in \cite{Romatschke:2007mq}. The shear viscosity in the partonic phase used here is twice the KSS limit \cite{Kovtun:2004de}. A bulk viscosity was also applied that peaks in the mixed phase, qualitatively consistent with both phenomenological models and with lattice extrapolations \cite{Paech:2006st,Karsch:2007jc}, but didn't lead to strong modifications to the source sizes. More details describing how bulk and shear viscosity are implemented can be found in \cite{Pratt:2008sz}.

Finally, we apply an improved treatment of the $\pi-\pi$ interaction used in calculating the relative wave function in the Koonin formula \cite{Lisa:2005dd,Koonin:1977fh} used to generate correlation functions,
\begin{eqnarray}
C({\bf k}_t,{\bf q})&=&\int d^3r ~S({\bf k}_t,{\bf r}) 
\left|\phi({\bf q},{\bf r}\right|^2\\
\nonumber
S({\bf k_t},{\bf r})&\equiv&\\
\nonumber
&&\hspace*{-30pt}\lim_{t'\rightarrow\infty}\frac{\int d^3r_1d^3r_2~f({\bf k}_t',{\bf r}_1',t')
f({\bf k}_t',{\bf r}_2',t')\delta({\bf r}-{\bf r_1}'+{\bf r_2}')}
{\int d^3r_1d^3r_2~f({\bf k}_t,{\bf r}_1,t)
f({\bf k}_t,{\bf r}_2,t)}.
\end{eqnarray}
Here, $f({\bf k}_t,{\bf r},t)$ represents the phase space density for pions of the relevant momentum, and the source function $S({bf k}_t,{\bf r})$ provides the normalized probability for two particles of the same momentum, ${\bf k}_t$, to be asymptotically separated by ${\bf r}$ in the pair rest frame (The primes refer to quantities in that frame). For the calculations presented thus far, a simple symmetrized plane wave for was used for the relative wavefunction, $|\phi|^2=1+\cos({\bf q}\cdot{\bf r})$,  which makes calculation of the correlation functions simple. More importantly, the correlation functions for Gaussian sources are then Gaussians in relative momentum, which makes the search for the best fit easy. If the sources were truly Gaussian, the radii extracted from this method would match those from a procedure using the full relative wave function including the strong and Coulomb interaction between pions. For the final calculation portrayed in Fig. \ref{fig:Rosl_everything} more realistic correlation functions were generated using the full wave function. These correlations were then  fit to Gaussian sources using the Bowler-Sinyukov method \cite{Bowler:1991vx,Sinyukov:1998fc} used in experimental analyses to mimic Coulomb effects. This results in a slightly improved fit to the experimental source sizes in Fig. \ref{fig:Rosl_everything}.

After the four improvements listed above, the model now reproduces the experimental source sizes to better than 10\% for all radii, which would seem to be within the systematic error quoted by the experiments, and close to the systematic uncertainty associated with the basic phenomenology. Thus, no single effect accounted for more than half of the HBT puzzle. Instead, the explanation amounted to a conspiracy of several effects, each of which increase the explosivity, and pushed the source sizes toward the data. Another modification, which will not be illustrated here, is to change the shape of the initial energy density profile. More compact profiles make the collision more explosive \cite{Broniowski:2008vp}. The profile applied here is that of the wounded nucleon model \cite{Kolb:2000sd}, but color-glass profiles

\begin{figure}
\centerline{\includegraphics[width=0.55\textwidth]{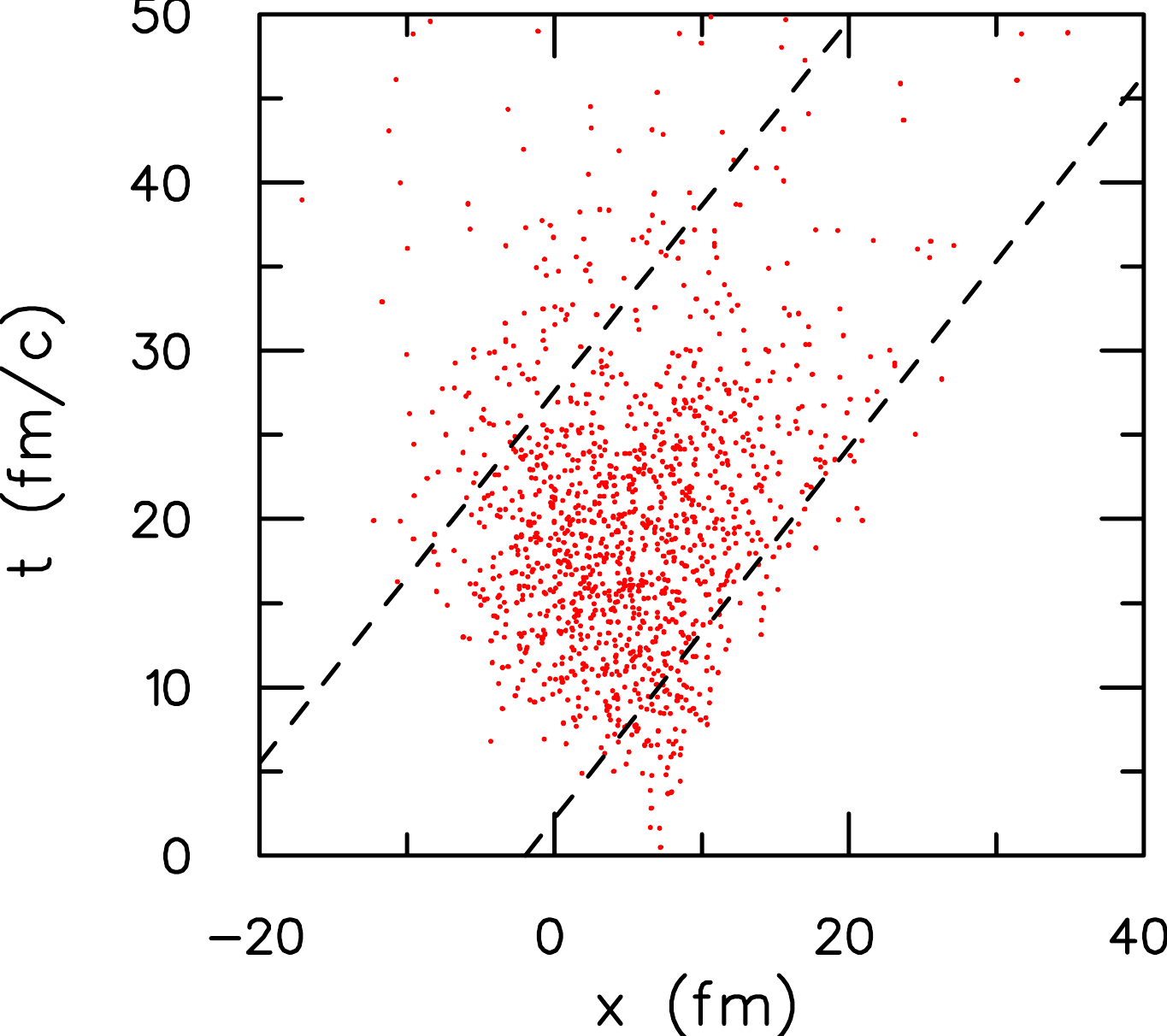}}
\caption{\label{fig:xyzt}
Final emission positions and times for particles with transverse momentum of 300 MeV/$c$ along the $x$ axis.  Emission has a positive correlation between position and time, though lags behind the slope of the velocity (illustrated by dashed lines). Due to the positive $x-t$ correlation, emissions of longer duration can still result in phase space clouds that are compact along the outward direction, with $R_{\rm out}/R_{\rm side}\approx 1$.}
\end{figure}

The second half of the HBT puzzle concerned the breakup densities inferred from the parameterized models \cite{Retiere:2003kf,Kisiel:2006is}. Over a thousand hadrons are produced in a central collision. For a blast-wave femtoscopic fitting exercises typically suggested radii near 12 fm, and a breakup time near 9 fm/c. The breakup volume would then be $\pi R^2\tau$, or $\approx 4000$ fm$^3$, corresponding to densities near 0.25 hadrons per fm$^3$, almost twice nuclear density. Given that hadronic cross sections tend to be $\sim 30$ mb, or 3 fm$^{2}$, the mean free path would be below 2 fm, which is far below the overall system size. The solution to this puzzle can be understood by viewing Fig. \ref{fig:xyzt}, which displays points sampling the final emission points from they hyrodynamic/cascade model for pions whose asymptotic momenta is 300 MeV/$c$ pointed along the $x$ axis. The $x$ and $t$ coordinates are clearly positively correlated, as expected for emission from an expanding surface, and are similar to what was seen in AMPT \cite{Lin:2002gc}. Since the source sizes characterize the shape of the outgoing phase space distribution for particles of this particular velocity, this correlation allows later produce pions to be produced in proximity to their earlier produced counterparts. If the particle emission had been confined to the neighborhood of a line with the slope of the velocity (see dashed lines), the emission duration would have been irrelevant, and $R_{\rm out}/R_{\rm side}$ would have been even lower than found here. The positive $x-t$ correlation allows the $R_{\rm out}/R_{\rm side}$ ratio to be near unity, even though the average emission time was near 20 fm/$c$. The parameterized forms do not properly account for this correlation, hence they provide a misleading picture of the breakup process.

The success of the fits in Fig. \ref{fig:Rosl_everything} are encouraging, but they immediately raise the question of whether the changes to the model destroyed the ability to fit spectra. To that end, Table \ref{table:meanpt} compares the mean $p_t$ values for the final model to experimental values from STAR and PHENIX. For protons, pions and kaons, the models value lied between the values reported by the two collaborations. Given that the previous hydrodynamic treatments also fit these values, it is surprising that the modifications presented here, all of which made the collision more explosive, did not result in more explosive spectra. In particular, we had expected the mean $p_t$ for protons to be over-predicted. Some of this may be due to the addition of bulk viscosity, which had little affect on HBT, but did lower the mean $p_t$. It is also possible, this agreement will disappear once the spectra are compared more carefully. In particular, it should be mentioned that the mean $p_t$ values from the model included products of weak decays. Filtering such products in a manner consistent with experiment might destroy the agreement.

Along with spectra and femtoscopic correlations, the third class of hadronic bulk observables are large-scale correlations, which are associated with collective flow. The elliptic-flow variable $v_2$ should be enhanced by the inclusion of pre-equilibrium flow, and by using a stiffer equation of state. Shear viscosity would be expected to lower the $v_2$, so it would not be surprising if these modifications resulted in little net change. Unfortunately, the hydrodynamic code used here assume radial symmetry, which precludes elliptic flow analysis. Nonetheless, the result found here provides significant hope that the entirety of soft hadronic observables can be reproduced with a single model. Doing so would represent a significant milestone for the field. This would be no means validate such a model, or its parameters, but it would demonstrate that a rigorous statistical assault on the data would result in a non-null region.

\begin{table}
\centerline{\begin{tabular}{|r|c|c|c|}\hline
	&$\pi^{(+,0,-)}$	&$K^{(+,-)}$	&$p,n,\bar{p},\bar{n}$\\ \hline\hline
STAR\cite{Abelev:2007rw}& $422\pm 22$	& $719\pm 74$	& $1100\pm 110$	\\ \hline
PHENIX\cite{Adler:2003cb}& $453\pm 33$	& $674\pm 78$	& $954\pm 85$	\\ \hline
Hydro+Cascade	& 433	& 714 	& 1027	\\ \hline
\end{tabular}}
\caption{\label{table:meanpt}
The mean $\langle p_t\rangle$ in MeV/$c$ for central collisions for pions, kaons and protons. Only charged species were used in the PHENIX analysis, and only negative hadrons were used for STAR. Since the original energy density was chosen to fit the multiplicity, the success seen here in reproducing the mean $p_t$ suggests that the spectral shapes are probably well described.}
\end{table}

\end{document}